\documentclass[mathleft
]{an}
\usepackage{graphicx}
\usepackage{times}
\begin{document}

\Pagespan{199}{202}
\Yearpublication{2009}%
\Yearsubmission{2008}%
\Month{99}%
\Volume{330}%
\Issue{2/3}%
\DOI{10.1002/asna.200811155}%

\title{A sample of GHz-peaked spectrum sources selected at RATAN--600: spectral and variability properties}

\author{K.~V.~Sokolovsky\inst{1,2}\fnmsep\thanks{Corresponding author:
  \email{ksokolov@mpifr-bonn.mpg.de}\newline}
Y.~Y.~Kovalev\inst{1,2}, Yu.~A.~Kovalev\inst{2},
N.~A.~Nizhelskiy\inst{3}, G.~V.~Zhekanis\inst{3}
}
\titlerunning{RATAN--600 GPS sample: variability and spectral properties}
\authorrunning{Sokolovsky et al.}
\institute{Max-Planck-Institute f\"ur Radioastronomie,
Auf dem H\"ugel 69, 53121 Bonn, Germany
\and
Astro Space Center of Lebedev Physical Institute,
Profsoyuznaya 84/32, 117997 Moscow, Russia
\and
Special Astrophysical Observatory RAS, Nizhnij Arkhyz, 369167 Russia}

\received{2008 Dec 8}
\accepted{2008 Dec 18}
\publonline{2009 Feb 15}


\abstract{%
We describe a new sample of 226 GPS (GHz-Peaked Spectrum) source
candidates selected using simultaneous 1--22~GHz multi-frequency
observations with the RATAN--600 radio telescope. Sixty objects
in our sample are identified as GPS source candidates for the first
time. The candidates were selected on the basis of their
broad-band radio spectra only. We discuss the spectral and variability
properties of selected objects of different optical classes.
}

\maketitle

\section{Introduction}

The first samples of GHz-Peaked Spectrum sources (e.g., Gopal-Krishna,
Patnaik \& Steppe 1983) were selected solely on the basis of radio
spectra properties. Later samples, like the one presented by Snellen et
al.~(2002) were considering other factors such as optical counterpart
type during the selection process. The very meaning of the word ``GPS
source'' has shifted from the description of spectral shape to something
close to ``probable young radio source'' which are believed to be found
among sources with convex spectra.

In this work we use the ``old fashioned'' spectral-only approach to select one
of the largest samples of peaked spectrum sources to date. Special care is
taken to keep the sample as free as possible from sources which change their overall
spectral shape due to variability. We investigate the observed properties of
sources from the sample to test the idea that sources with peaked spectra are
a mixture of intrinsically different types of AGN (e.g., Stanghellini 2003).

\section{Observational data and sample selection}

To select GPS source candidates we use 1--22~GHz multi-frequency data
from the RATAN--600 telescope of the Special Astrophysical Observatory,
Russian Academy of Sciences. It is a
576~m diameter ring radio telescope situated at stanitsa Zelenchukskaya,
Karachay-Cherkessia, Russia. The telescope is mostly operated in
transit mode. 
Emission from a radio source is measured as it crosses the feed beams of the
broad-band receivers operating at 1, 2.3,
3.9 (or 4.8), 7.7, 11 and 22 GHz. This allows us to obtain the source spectra
over a few minutes only. The exact time it takes the source to
cross all beams (and hence the integration time) is a function of
the source declination. Details of the method used for the observations and data
processing can be found in Kovalev et al.~(1999).

We use spectra of 4047 sources observed with
RATAN from 1997
to 2006 during multifrequency monitoring and survey campaigns conducted
by Kovalev et al.~(1999, 2000, 2002). The observed source list includes
a complete sample of sources with $\delta > -30$ and a total VLBI flux
density at 8~GHz greater than 200~mJy (Kovalev et al.~2007). To improve
the frequency coverage, we combine our RATAN--600 data with previously
published measurements collected by the CATS database (Verkhodanov et al.~2005).

\begin{figure*}[t]
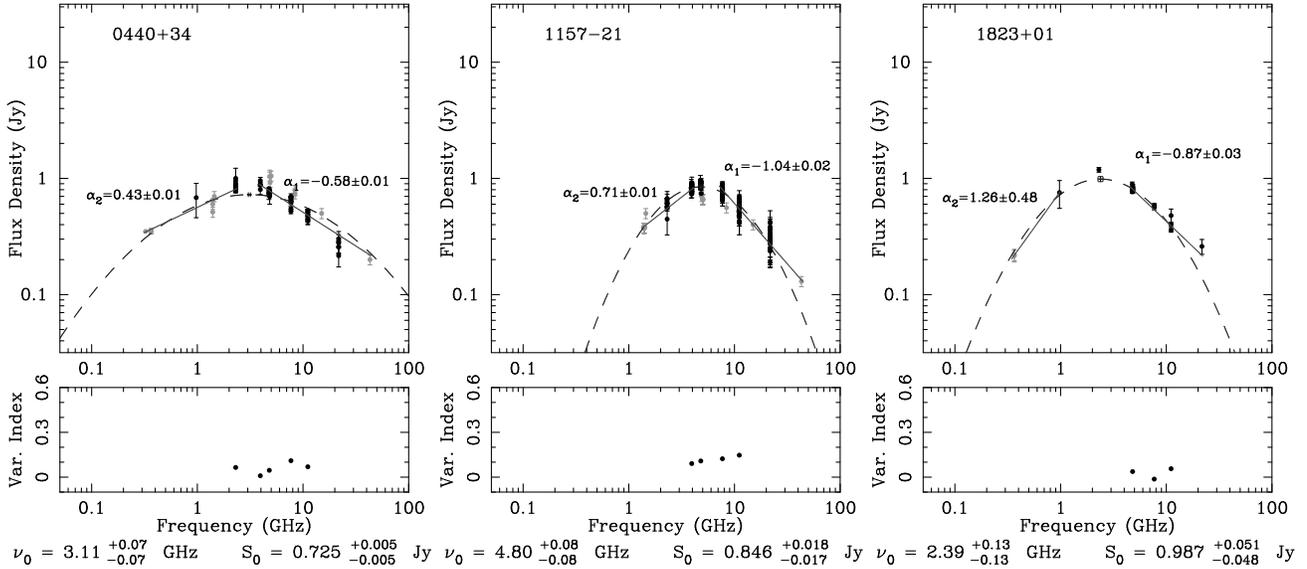

  \centering
  \includegraphics[trim=0cm 0cm 0.8cm 1cm,width=0.33\textwidth]{sokolovsky_fig1_0440p34s.ps}
  \includegraphics[trim=0cm 0cm 0.8cm 1cm,width=0.33\textwidth]{sokolovsky_fig1_1157m21.ps}
  \includegraphics[trim=0cm 0cm 0.8cm 1cm,width=0.33\textwidth]{sokolovsky_fig1_1823p0146.ps}
\caption{\label{fig:far-spect}
Radio spectra of 3 out of 60 newly identified GPS source candidates.
Variability index as a function of frequency is presented
in the panels under the spectra plots. Black points 
correspond to RATAN--600 data, grey points represent data collected
from the literature.}
\end{figure*}

\begin{figure}[th]
 \center{\includegraphics[angle=270,width=0.48\textwidth]{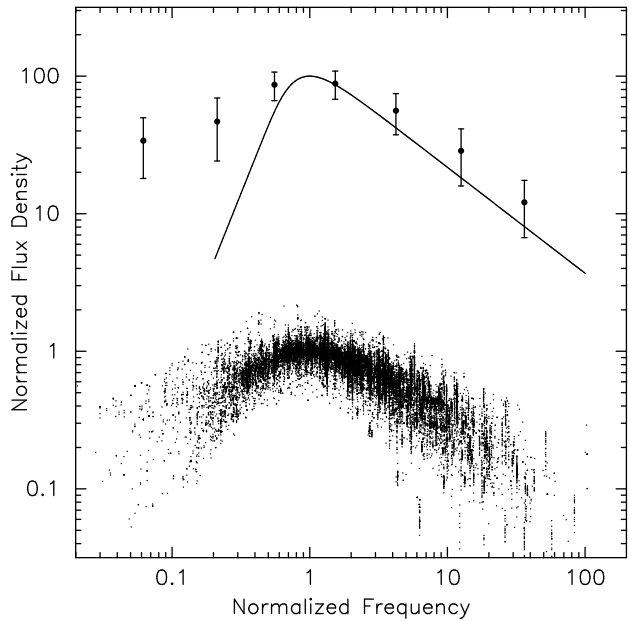}}
 \caption{ \label{fig:syntetic_synchro}
 Combined spectra of all selected 226 GPS candidates normalized by
 peak flux density and peak frequency values.
 Filled circles represent the observed spectra binned and shifted along
 the vertical axis. The solid line is a normalized and shifted
 theoretical spectrum of a homogeneous synchrotron emitting cloud with
 self-absorption and power index $\gamma=2.54$ of electrons energy
 distribution.
}
\end{figure}

We inspected all of the collected broad-band spectra to select sources
which show a prominent spectral peak at all epochs for which we have
RATAN data (supplemented by the CATS database). To derive the peak flux
density ($S_{\nu_0}$) and peak frequency ($\nu_0$), we followed O'Dea et
al.~(1990), Snellen et al.~(2002) and Edwards \& Tingay~(2004), and
fitted each source spectrum in the log-log scale with a parabolic
function $\log S_\nu = a (\log \nu)^2 + b \log \nu + c$, were
$S_\nu$ is the flux density at frequency $\nu$, $a$, $b$, and $c$ are
coefficients determined using a least square regression method. Despite
having no physical meaning, this simple fit can represent observations
reasonably well. The spectral indexes\footnote{We use the following
spectral index $\alpha$ definition: $S_{\nu} \sim \nu^{\alpha}$.
$\alpha_1$ is a spectral index above the peak frequency $\nu_0$,
$\alpha_2$ -- below the peak.} were determined by fitting the high-
and low-frequency regions of the spectra with a linear function in the
log-log scale. We have used similar criteria to those suggested by
de~Vries, Barthel \& O'Dea~(1997) to select the final list of GPS
candidates: {\em{}(i)}~source spectra show a defined peak at a frequency
above $500$~MHz, {\em{}(ii)}~the difference between the high- and
low-frequency spectral indexes is greater than $0.6$. We use the
traditional $500$~MHz boundary between GPS and Compact Steep Spectrum
(CSS) sources, despite there is no clear physical boundary between
them, because we do not have our own observations at frequencies below
1~GHz.

We have selected a sample of 226 GPS source candidates, 60 objects from
the sample have never been reported before as GPS source candidates
(see examples in Fig.~\ref{fig:far-spect}).
Optical identifications and redshift information were extracted from the
V{\'e}ron-Cetty \& V{\'e}ron~(2006) catalog and NED (NASA/IPAC
Extragalactic Database). Spectroscopic redshifts were found for 128
sources, 39 sources are identified with galaxies, 98 are identified with
quasars, three sources (PKS 0637$-$337, PKS 1300$-$105, and
PKS 1519$-$273) are BL~Lacertae objects. For 86 sources no 
optical identification was found (empty fields).

\section{Spectral properties}

An ``average'' GPS spectrum in our sample is characterized by the peak
frequency $\nu_{0} = 2.4$~GHz and flux density $S_{\nu_{0}} = 0.8$~Jy. The
spectral indexes above and below the turnover frequency are $\alpha_{1} =
-0.77$ and $\alpha_{2} = 0.66$. 

To provide a quick impression of the spectra selected, we present
Fig.~\ref{fig:syntetic_synchro}. An average spectrum of a GPS source was
constructed by combining the individual source spectra normalized by their
peak frequency and flux density. This combined spectrum is compared to a
theoretical spectrum of a homogeneous synchrotron emitting cloud with
a spectral index above the peak equal to the
median value for our sample. 

As can be seen from
Fig.~\ref{fig:syntetic_synchro}, the observed spectra of GPS sources are
wider and do not reach $\alpha_2=5/2$ which is predicted by synchrotron
self-absorption. In fact, no single source in our sample approaches this
value. The most inverted $\alpha_2=1.76\pm0.03$ is found in the known GPS quasar
0457+024. Among the 10 sources with the most inverted spectra there are
five quasars, one radio galaxy and four empty fields.
The absence of observed $\alpha_2 \sim 5/2$ means that the radiation of GPS sources can not be described
assuming one homogeneous synchrotron self-absorbed cloud. There should
be a significant non-homogeneity in the properties across the volume of
the radio--emitting region.

The broad-band radio spectra of GPS galaxies and quasars look very much
alike. We found no significant differences in the low-frequency spectral
index distribution between subgroups of our sample: galaxies, quasars
and unidentified radio sources. This may be an indication of a common
mechanism responsible for the absorption at low frequencies.
The median value of high-frequency spectral index is
$\alpha^\mathrm{gal}_1=-0.80$ for galaxies and
$\alpha^\mathrm{qso}_1=-0.71$ for quasars. The Kolmogorov-Smirnov (KS)
test has also confirmed that the distribution for galaxies and quasars
are not different significantly.

\begin{figure}[t]
 \centerline{\includegraphics[width=0.5\textwidth]{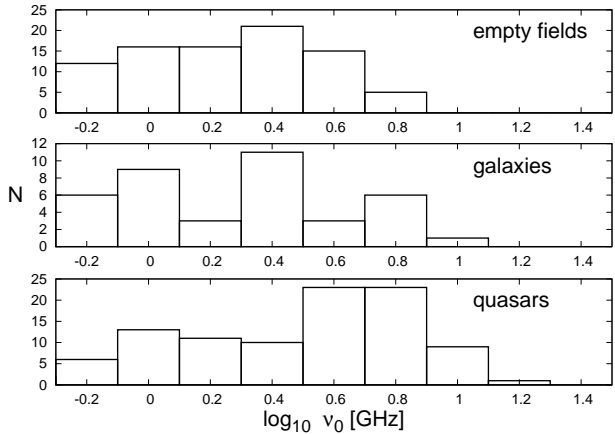}}
 \caption{Peak frequency distribution for GPS
 quasars, galaxies and unidentified radio sources (empty fields).}
 \label{fig:pfreq}
\end{figure}

No difference was found in the observed peak flux density between
galaxies and quasars. However, when it comes to the peak frequency
distribution (Fig.~\ref{fig:pfreq}), the KS test shows that there is
only a 10 per~cent chance that the observed peak frequency distributions
of galaxies and quasars are drawn from the same parent distribution. If
we consider the peak frequency in the source frame, this probability
drops to 0.01 per~cent. Galaxies in our sample are characterized by a
lower intrinsic peak frequency (median value 3~GHz) than quasars
(8.8~GHz). This is expected since quasars are found, in general, at
higher redshifts.

\section{Variability}

For the 108 sources which have observations with RATAN--600 at three
or more epochs, a variability study was carried out. We concentrate on the
variability properties at 11~GHz which corresponds to the
optically thin part of the synchrotron spectrum for the majority of the
sources. In addition, this band is almost RFI free, has high
sensitivity and is not affected significantly by weather. Sources with
a peak frequency around or above 11~GHz are excluded from the analysis in
order to probe variability at a frequency with no self-absorption.

To characterize the variability amplitude of sources we use the variability index (e.g., used by
Aller, Aller \& Hughes~1992) applying the modified robust form:
$$ v =
\frac{(S_{i}-\sigma_{i})_\mathrm{max}-(S_{i}+\sigma_{i})_\mathrm{min}}{(S_{i}-\sigma_{i})_\mathrm{max}+(S_{i}+\sigma_{i})_\mathrm{min}}\,
$$
$S_{i}$ and $\sigma_{i}$ are the source flux density and RMS error
measured at $i$'th epoch, max/min indexes correspond to the
maximum/minimum value among all epochs. Note, that $v$ can be negative if
the estimated error $\sigma$ is greater than the observed scatter of the
data. 
The relative accuracy of flux density measurements as well as the number of
observations influences the value of $v$. We have modeled the
variability index distribution for a sample of non-variable sources
which have fluxes that are measured with a typical accuracy of the RATAN measurements 
(assuming Gaussian noise). In this modeling we have used the actual
distribution of the number of observations. We conclude that sources
with $v \ge 0.15$ can be considered as significantly variable.

\begin{figure}[t]
 \centerline{\includegraphics[width=0.5\textwidth]{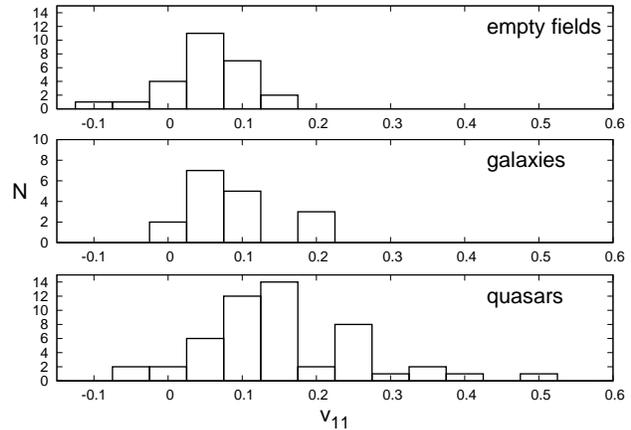}}
 \caption{\label{fig:v11}
 Distribution of variability index $v_{11}$ at 11~GHz for GPS
 quasars, galaxies and unidentified radio sources (empty fields)
 in our sample. 
 }
\end{figure}

The distribution of $v_{11}$ is found to be different for
galaxies and quasars
(Fig.~\ref{fig:v11}). There are many variable sources among quasars. 
The statistics for galaxies is suffering from the small number of objects, but
it can be clearly divided into two groups: the main distribution which is
consistent with non-variable or weakly variable sources,
with variations under 9--10 per~cent.
The second group consist of three strongly variable
galaxies: PKS~0500$+$019, note that this object is listed as a quasar
by Becker, White \& Edwards~(1991); TXS~1404$+$286 (OQ~208,
Mkn~668), its variability at a frequency above the peak was first
reported by Stanghellini et al.~(1997); and B2~1600$+$33.

\begin{figure}[t]
 \centerline{\includegraphics[width=0.5\textwidth]{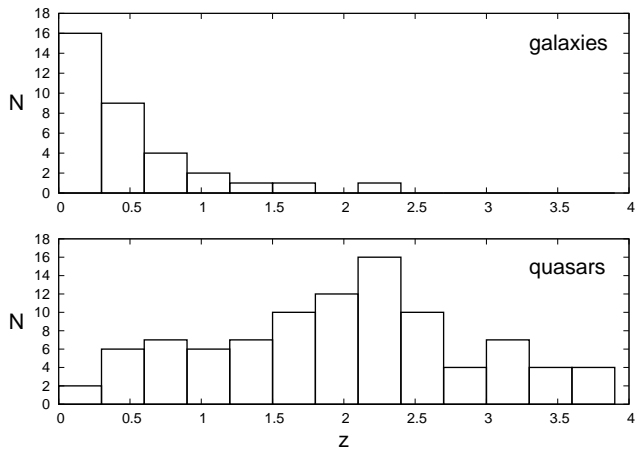}}
 \caption{Redshift distribution of GPS galaxies and GPS quasars from the
 sample.}
 \label{fig:z}
\end{figure}

A KS test gives a 63 per~cent probability that galaxies (with the exception
of three clearly variable cases) and ``empty fields'' are drawn from the
same distribution of variability index at 11~GHz. The same
probability for galaxies and quasars is neglectable ($< 0.01$~per~cent). This
implies that most of the optically unidentified GPS sources might be
faint distant galaxies and not quasars.
The similarity of the observed peak frequency distribution for galaxies and
``empty fields'' (Fig.~\ref{fig:pfreq}) supports this conclusion.
This may suggest, that the observed difference in redshift
distributions of GPS quasars and galaxies (Fig.~\ref{fig:z}) can not be used as an argument in
support of the idea that these two types of GPS sources are intrinsically
different. If there is a large
number of distant GPS galaxies which lack an optical identification
that are also
optically faint, then the difference in the redshift distribution may be
attributed to a selection effect.

\section{Summary}

We have selected a sample of 226 GPS source candidates from 
multi-frequency, multi-epoch observations conducted at the 
RATAN--600 radio telescope, supplemented by the literature; 60
objects from the sample are reported as GPS candidates for the first
time.

Derived parameters of the convex spectra
do not allow us to distinguish between GPS galaxies and quasars. GPS
quasars tend to show significantly stronger variability than GPS galaxies,
which is in agreement with previously published results for both GPS
(e.g., O'Dea~1998) and high frequency peaked (Tinti et al.~2005)
sources. Most GPS galaxies and ``empty fields'' show no significant
variability with three noticeable exceptions: PKS~0500$+$019,
TXS~1404$+$28, and B2~1600$+$33.
The low variability of GPS sources with no optical identification suggests
that a significant fraction of them might be optically faint and,
possibly, distant GPS galaxies.
GPS galaxies and GPS quasars have significantly different variability
properties and intrinsic peak frequencies. However, no clear boundary
can be drawn between them on the basis of single-dish radio observations
because GPS galaxies and GPS quasars strongly overlap in their
observed properties.

We plan to extend our study of the selected GPS sample to parsec
scales by analyzing simultaneous 2 and 8 GHz VLBA observations from the 
VLBA Calibrator survey (Beasley et al.~2002, Fomalont et al.~2003,
Petrov et al.~2004, 2005, 2008, Kovalev et al.~2007).

\acknowledgements
K.~Sokolovsky is supported by the International Max Planck
Research School (IMPRS) for Radio and Infrared Astronomy, his participation
in the $4^{th}$ CSS/GPS workshop was partly supported by funding from the
European Community's sixth Framework Programme under 
RadioNet R113CT 2003 5058187.
Y. Y. Kovalev is a Research Fellow of the Alexander von Humboldt
Foundation. RATAN--600 observations are partly supported by the
Russian Foundation for Basic Research (projects 01-02-16812,
05-02-17377, 08-02-00545). 
The authors made use of the database CATS (Verkhodanov et al.~2005)
of the Special Astrophysical Observatory.
This research has made use of the NASA/IPAC Extragalactic Database (NED)
which is operated by the Jet Propulsion Laboratory, California Institute
of Technology, under contract with the National Aeronautics and Space
Administration. This research has made use of NASA's Astrophysics Data
System. The authors are grateful to John McKean and to the two anonymous 
referees for their constructive comments which helped to improve the
manuscript.



\begin{thebibliography}{}
  \bibitem{} Aller, M.F., Aller, H.D., Hughes, P.A.: 1992, ApJ 399, 16
  \bibitem{} Beasley, A.J., et al.: 2002, ApJS 141, 13
  \bibitem{} Becker, R.H., White, R.L., Edwards, A.L.: 1991, ApJS 75, 1
  \bibitem{} de Vries, W.H., Barthel, P.D., O'Dea, C.P.: 1997, A\&A 321, 105
  \bibitem{} Edwards, P.G. and Tingay, S.J.: 2004, A\&A 424, 91
  \bibitem{} Fomalont, E.B., et al.: 2003, AJ 126, 2562
  \bibitem{} Gopal-Krishna, Patnaik, A.R., Steppe, H.: 1983, A\&A 123, 107
  \bibitem{} Kovalev, Yu.A., Kovalev, Y.Y., \& Nizhelsky, N.A.: 2000, PASJ 52, 1027
  \bibitem{} Kovalev, Y.Y., et al.: 1999, A\&AS 139, 545
  \bibitem{} Kovalev, Y.Y., Kovalev, Y.A., Nizhelsky, N.A., 
Bogdantsov, A.B.: 2002, PASA 19, 83
  \bibitem{} Kovalev, Y.Y., Petrov, L., Fomalont, E.B., \& Gordon, D.: 2007, AJ 133, 1236
  \bibitem{} O'Dea, C.P., Baum, S.A., Stanghellini, C., 
Morris, G.B., Patnaik, A.R., Gopal-Krishna: 1990, A\&AS 84, 549
  \bibitem{} O'Dea, C.P., 1998, PASP 110, 493
  \bibitem{} Petrov, L., Kovalev, Y.Y., Fomalont, E.B., \& Gordon, D.:
2008 AJ 136, 580
  \bibitem{} Petrov, L., Kovalev, Y.Y., Fomalont, E.B., \& Gordon, D.:
2006, AJ 131, 1872
  \bibitem{} Petrov, L., Kovalev, Y.Y., Fomalont, E.B., \& Gordon, D.:
2005, AJ 129, 1163
  \bibitem{} Snellen, I.A.G., Lehnert, M.D., Bremer, M.N., Schilizzi, R.T.: 2002, MNRAS 337, 981
  \bibitem{} Stanghellini, C., Bondi, M., Dallacasa, D., O'Dea, C.P., Baum,  
S.A., Fanti, R., Fanti, C.: 1997, A\&A 318, 376
  \bibitem{} Stanghellini, C., O'Dea, C.P., Dallacasa, D.,
Baum, S.A., Fanti, R., Fanti, C.: 1998, A\&AS 131, 303
  \bibitem{} Stanghellini, C.: 2003, PASA 20, 118
  \bibitem{} Tinti, S., Dallacasa, D., de Zotti, G., Celotti, A., 
Stanghellini, C.: 2005, A\&A 432, 31
  \bibitem{CATS} Verkhodanov O.V., Trushkin S.A., Andernach H., \& Chernenkov V.N.:
2005, Bull.~SAO 58, 118; arXiv:0705.2959
  \bibitem{} {V{\'e}ron-Cetty}, M.-P. and {V{\'e}ron}, P.:
2006, A\&A 455, 773
\end{thebibliography}
\end{document}